\documentclass[aps,pra,nofootinbib,twocolumn,longbibliography]{revtex4-1}

\usepackage{amsfonts}
\usepackage{amsmath}
\usepackage{graphicx}
\usepackage{amssymb}
\usepackage{times}
\usepackage{color}
\usepackage[colorlinks,breaklinks]{hyperref}

\newcommand{\ket}[1]{|#1\rangle}
\newcommand{\bra}[1]{\langle#1|}

\newcommand{\fket}[1]{|#1\!\succ}

\newcounter{myenumi}
\begin{document}

\title{Perhaps they are everywhere? Undetectable distributed quantum computation and communication for alien civilizations can be established using thermal light from stars}
\author{Terry Rudolph}
\affiliation{PsiQuantum \& Imperial College London}

\begin{abstract}

We show that free-space diffraction of photons distributes highly useful entanglement:  the receivers of the propagated modes can do a distributed quantum computation using only linear optics and photon counting. The distributed computation requires classical communication between receivers, however, similar to standard measurement-based computation, that communication is of purely random outcomes and so can be indistinguishable from noise. The speculation in the title arises from the further observation that the natural way for a circumspect civilization to hide their photonic entanglement distribution is to use the thermal light already being emitted from the various stars they visit. This requires them knowing the number of photons in the modes they have chosen to use, and as such they would need to perform a quantum non-demolition measurement of photon number. Because the thermal light they are measuring is diagonal in the number basis even this process can be rendered in principle indiscernible to those of us excluded from the conversation.   
 \end{abstract}
 
\maketitle
\pagestyle{plain}


\section{Introduction}

We will establish that the free space diffraction of single photons deterministically both generates and shares entanglement which is powerful enough to enable distributed universal quantum computation. The computation itself requires the receivers to use only linear optics (passive interferometers) and photon number counting, and to be able to classically communicate. 

More precisely, consider the state of a single photon spread uniformly across $K$ modes, i.e. a kind of ``single photon W-state'':
\begin{align}\label{eq:Wstate}
\ket{W_K}&=\frac{1}{\sqrt{K}}\sum_{j=1}^Ke^{i\theta_j}\fket{1_j} \\
&=\frac{1}{\sqrt{K}}(e^{i\theta_1}\fket{10..0}\!+e^{i\theta_2}\fket{01..0}\!+\ldots e^{i\theta_K}\fket{00..1}). \nonumber
\end{align}
If the modes in question are spatially distinct then $K$ separated receiving parties can each accept one of the modes. By taking $N$ independent photons distributed in this manner  (i.e. preparing $\ket{W_K}^{\otimes N})$ each receiver will hold $N$ modes. Choosing $K>>N$ we can ensure that every receiver would detect at most one photon should they happen to measure all modes in their lab (most receivers would discover only vacuum).

Our main observation will be that entanglement generated and distributed in this simple and minimalistic way is sufficient for the $K$ receivers to efficiently run a universal quantum computation. Note that no interaction or interference between the photons plays any part in the distribution.
Getting to the result will involve the somewhat absurd twist of encoding first-quantization into a second-quantized state, and we are forced to annoyingly refer to such as a \emph{third quantized} state.

Photons can propagate billions of light years and retain significant quantum coherence (e.g. we receive polarized light from Lyman-$\alpha$ blob 1). One consequence is therefore that a sufficiently advanced civilization can perform quantum non-demolition measurements of photon number on suitable modes of light being emitted from stars, in such a way that useful large-scale entanglement is distributed by the subsequent free-space propagation of that light through the universe\footnote{Paranoid aliens presumably use only Hawking radiation. Particularly paranoid ones use only one mode per black hole. It is not necessary, however, that the specific modes being used are secret. Even if they are known, one convex decomposition of a thermal density matrix $\rho_{therm}$ is into an incoherent mixture of coherent states $\ket{\alpha}$, i.e. $\rho_{therm}=\int d^2\alpha P(|\alpha|^2)\ket{\alpha}\bra{\alpha}$. Coherent states evolve into product states under diffractive propagation/linear optics and thus the evolved density matrix is always separable to an eavesdropper without knowledge of the photon number. That is, access to the entanglement requires knowledge of each specific mode's occupation number.  This gives an additional layer of obfuscation over the fundamental quantum information-theoretic security one gets from protocols like quantum key distribution.}. The thermal density matrix is unperturbed by the QND measurement, as it is diagonal in this basis. Using this entanglement (for quantum computing, or quantum secret sharing, or quantum data hiding, or quantum key distribution - or whatever you think aliens are up to) relies on classical communication of measurement outcomes between the receiving nodes. One may hope, therefore, that we could detect the presence of said aliens by looking for that communication. Unfortunately for those of us who would like to listen in, similar to a one-way computation on qubits \cite{Raussendorf2001}, the local measurement outcomes - and therefore messages that need to be communicated - are themselves indistinguishable from thermal noise.

The upshot is that when we look to the stars and see only thermal radiation we typically conclude the universe is empty. But perhaps, riding in the correlations of that radiation, the universe is actually bathed in alien chatter and other forms of distributed quantum information processing. Unfortunately this is all fundamentally hidden from us if quantum theory is correct. And if it isn't correct, then presumably the aliens know that and so are not using this method. It seems, therefore, that the only way to test this hypothesis is to wait for them to drop by (again?) and let us know which case pertains.  

\section{Review: Linear optical manipulation of modes}\label{sec:LOmanipulation}

We will be concerned only with linear optical manipulation of photons - that is, evolution that can be performed by an interferometer. This ``free field'' evolution (of which free-space diffraction is a case) cannot change photon number - it commutes with the free Hamiltonian - and so does not result in photon interactions per se. But it does allow for non-trivial manipulation of modes, quantifiable via a unitary operator acting on the creation operators $a_i^\dagger\rightarrow \sum_j U_{ji}a_j^\dagger$. Considering modes rather than photons as the ``systems'' of relevance this manipulation is highly entangling although considerably constrained - within the subspace of states with a fixed number of photons not every state can be reached from every other state. 

When the subspace is that of only one photon in (say) $d$ modes then all states can be reached by an interferometer: an arbitrary unitary evolution within the $d$-dimensional subspace is possible by using an interferometer whose evolution of creation operators for the modes is given by the same unitary matrix. As such we can perform any evolution or measurement on a single qu\emph{d}it if we encode it using one photon in $d$ modes. 


A particularly cheap way to deterministically create bipartite entanglement is to create a single photon in superposition across two modes (eg using a 50:50 beamsplitter). The resultant state $\fket{1}\!\!\fket{0}\!\!+\fket{0}\!\!\fket{1}\,$ (curved brackets $\fket{\cdot}$ always denote second-quantized (Fock) states)  contains entanglement between two modes, not two particles, but it is standard to refer to this as single photon entanglement. This bipartite entanglement can be operationally useful\footnote{See e.g. \cite{vanenk1001} for a discussion of how to use it for protocols such as teleportation, as well as a discussion of historical controversies about whether such entanglement is ``genuine single photon entanglement'' or not.}.

It is natural to wonder then if the state $\ket{W_K}$ is ``even better'' as a source of entanglement in some quantifiable sense? Multipartite W-state entanglement of qubits is generally much less interesting and useful than (say) stabilizer multiqubit entanglement. However the comparison is risky, because we can easily and cleanly manipulate multiple photonic modes simultaneously using a linear-optical interferometer. From a qubit perspective this would amount to performing entangling operations with highly nontrivial interactions.

Below we will show the following: The $K$-partite entanglement given $N$ copies of a $\ket{W_K}$ state is universal for  quantum computing problems of size (number of qubits and gates) $N^c$, for some constant $c>0$, even if $K$ is so large in comparison to $N$ that no party ever detects more than one photon (e.g. taking $K=N^3$ would suffice).   

This should be contrasted to the standard type of things we expect from an architecture for  photonic quantum computing: we expect to need entangled stabilizer states, for generation of that entanglement to be non-determinisitic, we expect multiphoton interference (e.g. the ``HOM'' dip) to play an important role and so on. The protocol we present will be most similar to measurement based (or ``one-way'') quantum computing \cite{Raussendorf2001, Gross2007}. But there are significant differences, including the fact that our protocol will involve all parties doing a measurement - in fact exactly the same measurement - at every step, but most outcomes will be of high rank so that the initial entanglement is not immediately consumed. 

Many of the ingredients required to understand how to do universal quantum computing using $\ket{W_K}^{\otimes N}$ as the basic resource are easiest to introduce by considering using single photons for the simpler task of performing a Bell experiment.

\section{Warmup: Bell experiments using single photons spread over many modes}\label{sec:warmupbell}

With two copies of  $\fket{1}\!\fket{0}+\fket{0}\!\fket{1}$, giving one mode from each copy to Alice and the other to Bob, we have a state of the form:
\begin{align*}
\ket{W_2}^{\otimes 2}= &\fket{11}_A\fket{00}_B+\fket{00}_A\fket{11}_B\\+&\fket{10}_A\fket{01}_B+\fket{01}_A\fket{10}_B. 
 \end{align*}
As mentioned, within the subspace spanned by a single photon in $d$ modes there is no constraint imposed by the restriction to only linear optical evolution and measurement. From the latter two terms of the expression above we see that half the time Alice and Bob will each find only one photon in the two modes in their lab, and on these occasions they can therefore perform a standard qubit based Bell experiment. That is, in that part of the total wavefunction they hold a regular dual-rail encoded maximally entangled state $\ket{\Psi_{AB}^+}=(\fket{10}_A\fket{01}_B+\fket{01}_A\fket{10}_B)$.

From the perspective of testing local realism such an experiment is fine, but it is clearly inefficient because half the time one or the other party detects two photons, and data from those runs of the experiment lead to classical correlations and must be discarded. That is, with linear optics there is no way for them to perform measurements in the subspace spanned by $\fket{00}$ and $\fket{11}$.

If photons are more expensive than grad students, then one way to arbitrarily increase the efficiency of this kind of Bell experiment is to use many more parties. That is, we can spread the two photons over $K>>1$ modes. Consider expanding out the state $\ket{W_K}^{\otimes 2}$ and regrouping the modes so that $K$ different parties each hold 2 modes in their lab. For large enough $K$ the part of the wavefunction in which both photons end up at the same lab can be ignored. The remaining part of the wavefunction can be interpreted as ``a random pair of parties share a regular maximally entangled state'':
\begin{align*}
\ket{W_K}^{\otimes 2}\approx \ket{\Psi_{AB}^+}+\ket{\Psi_{AC}^+}+\ldots+\ket{\Psi_{BC}^+}+\ldots+\ket{\Psi_{YZ}^+}.
\end{align*}
where the parties not indicated in each term hold vacuum.

We now imagine every party performs a measurement as if they were holding one qubit of the maximally entangled state. On every run a random pair of parties will record an outcome, while $K-2$ of the parties will detect vacuum. From the perspective of testing local realism such an experiment (in the limit of large $K$) approaches unit efficiency (in terms of number of photons, not grad students, that get used up) because the inefficient cases where two photons ended up in one lab no longer occur.\footnote{There is a subtlety for this type of  Bell experiment. Namely, in a standard CHSH version of a Bell experiment the measurements performed by Alice and Bob are not identical. In our multi-party variation each of the parties do not know whether they are holding a photon until its too late (i.e. they have detected it), so how do they know whether to be Alice or to be Bob in the protocol? Choosing random measurements will, with high likelihood, suffice to violate some kind of Bell inequality \cite{randombell}, but it won't be maximal, and so more precious photons than necessary might need to be used. Fortunately there exists a two qubit state lying in the symmetric subspace such that maximal CHSH violation can be achieved by Alice and Bob both choosing from the exact same set of measurements, say Pauli $X$ and $Z$ basis projections. Moreover, this state can be reached from a standard Bell state $\ket{\Psi_{AB}^+}$ using identical single qubit rotations $U\otimes U$ on each qubit. Applied to the photonic protocol, we determine that as long as every party implements the single qubit unitary $U$ prior to making their random choice of $X$ or $Z$ measurement, maximal violation will be achieved. For the case of quantum computation, our primary focus here, this subtlety plays no role.}          

It is worth noting a couple of features of this protocol: 
\begin{itemize}
\item If every party applies a random phase rotation to their whole lab (i.e. every mode in their lab gets the same random phase) then they will end up sharing a uniformly mixed state 
$$\ket{\Psi_{AB}^+}\bra{\Psi_{AB}^+}+\ldots+\ket{\Psi_{YZ}^+}\bra{\Psi_{YZ}^+}
$$
over the Bell pairs, but this will not affect the efficiency of the protocol. Thus there is no need to keep the labs all phase  stable with respect to each other. Another way of saying this is that the phases $\theta_i$ in the definition of $\ket{W_K}$ do not need to be known, as long as they are the same in all $N$ copies of $\ket{W_K}^{\otimes N}$.
\item In some sense the two single photons involved never ``see'' each other. They are each prepared in the initial uniform superposition $\ket{W_K}$ completely independently, and they end up in different labs - no multiphoton terms such as $\fket{2}$ are relevant to the whole procedure. The only way in which they ever see each other is when they are interfered with a mode that the other photon could potentially have been in (but of course is certain to not actually be in!). The interference driving everything is more wavelike than particle-like; this extends to the model of full quantum computation below.
\item If the amplitudes in $\ket{W_K}$ were not uniform in magnitude they can be easily rebalanced if necessary using something akin to standard procrustean distillation.
\end{itemize}

\section{Reminder: First-quantized description of protocols involving linear optics and single photons}

Consider how we describe quantum-mechanically three single photons that happen to be in modes labelled $a$, $b$, $c$. In second quantization this state would be $\fket{1_a}\fket{1_b}\fket{1_c}$. In second quantization the systems are modes, and photons are internal level occupations of a mode. If, as is the case here, we never interact the photons (i.e. use only linear optics and measurements diagonal in photon number) anything we do will also have some satisfactory description in a first-quantized picture. The initial state of the photons would then be the symmetrized state $\ket{\sigma_{111}}$:
\begin{align}\label{sigma111}
\fket{1_a1_b1_c}\,\Leftrightarrow\ket{\sigma_{111}}=&\ket{a}\ket{b}\ket{c}+\ket{a}\ket{c}\ket{b}+\ket{b}\ket{a}\ket{c} \nonumber  \\
+&\ket{b}\ket{c}\ket{a}+\ket{c}\ket{a}\ket{b}+\ket{c}\ket{b}\ket{a}.
\end{align}
In this first-quantized description the systems are photons and the modes are internal level occupations of the photons. 

If we sent the three photons through an interferometer described by mode-transformation $U$ then the state in first quantization would evolve to $(U\otimes U \otimes U) \ket{\sigma_{111}}$. This is simpler than describing evolution through an interferometer in second quantization. However, if we then did photodetection on, say, mode $a$, the first-quantized description of that measurement would involve a joint (i.e. entangling) projector across all three systems (i.e. photons) and a messy collapse process to describe the state remaining in modes $b,c$. In second quantization the same measurement acts simply on a single system (i.e. mode).

\section{Getting to full quantum computation}\label{sec:gettingtofull}

In this section we will see that the generalization to $N$ photons of the protocol in Section~\ref{sec:warmupbell} actually empowers the $K$ parties to run a universal quantum computation using only local (linear optical) operations and classical communication. All the entanglement driving the quantum computation is being deterministically prepared and distributed via the uniformly spread out single photons. For simplicity $K$ will be chosen large enough that the probability of any party ever detecting two photons in their lab is negligible. 

Getting to the final result is going to involve a slightly ludicrous twist that can be pretty confusing. Fortunately it can be understood by considering the example of 3 photons. That is, we consider $N=3$ photons distributed as above between $K$ parties. Similar to the case above of the Bell experiment, we can expand the state over all possible triples of parties as follows:
\begin{align}\label{WK3}
\ket{W_K}^{\otimes 3}\approx \ket{\Sigma_{ABC}}+\ket{\Sigma_{ABD}}+\ldots+\ket{\Sigma_{XYZ}},
\end{align}
where the state of the three photons held by, say, the trio of parties $R$, $S$, $T$ is given by
 \begin{align}\label{Sigma3}
\ket{\Sigma_{RST}} 
  = &\frac{1}{\sqrt{3!}}(\fket{100}_R\fket{010}_S\fket{001}_T\nonumber\\ &+\fket{100}_R\fket{001}_S\fket{010}_T +\ldots\nonumber\\
   & +\fket{001}_R\fket{010}_S\fket{100}_T),
\end{align}
(the remaining parties hold vacuum).

Comparing $\ket{\sigma_{111}}$ of Eq.~(\ref{sigma111}) and $\ket{\Sigma_{RST}}$ we see an immediate correspondence. The state $\ket{\Sigma_{RST}}$, viewed in second quantization, encodes the maximally-symmetric, first-quantized state of three photons if each party $R$, $S$, $T$ interprets the single-photon qutrit they hold in the three modes in their lab appropriately. We will see, in fact, it can be used to simulate any protocol on three photons, by mimicking the first-quantized description of that protocol. 

This second-quantized-encoding-of-a-first-quantized-state, that we will term \emph{third-quantized}, has pitfalls for the unwary. Despite being a three photon state, $\ket{\Sigma_{RST}}$ is a state of three photons in nine modes, not three modes, and so its own first-quantized description would \emph{not} be of the form of Eq.~(\ref{sigma111}). It is also (mode-)entangled, whereas the state $\fket{111}$ that it is simulating is not. 

What we first want to understand now is how, if three parties happened to be given the entangled third-quantized state $\ket{\Sigma_{RST}}$ of Eq.~(\ref{Sigma3}), they could simulate any first-quantized description of an arbitrary photonic protocol one might perform on three single photons.

The case of how to simulate evolution - an interferometer - is easy: As mentioned in Section \ref{sec:LOmanipulation}, evolution of a qudit encoded via a single photon in $d$-modes can be fully simulated using only linear optics. So whatever evolution $U$ describes the interferometer the photons go through, parties $R$, $S$ and $T$ can independently implement it on their three modes, and obtain the $U\times U \otimes U$ evolution that mimics the first-quantized case.

Simulating the measurement seems considerably more problematic. Especially so, because we are aiming for a simulation where parties $R$, $S$ and $T$ would only classically communicate, and so performing a joint - i.e. entangling - measurement (to mimic a first quantization measurement) would be impossible without ancillary resources such as shared Bell pairs. This is where the following nice observation of Popescu \cite{popescuKLM} comes to the rescue: If nature allowed us to perform independent measurements on the systems (photons) of first quantization (it doesn't!) then it would not actually change the measurement statistics observed as long as all the measurements we did were identical. That is, a hypothetical independent measurement on the three photons, one where each photon underwent the same POVM - say $\{\ket{a}\bra{a},I-\ket{a}\bra{a}\}$ to model photodection in mode $a$ - would result in the same outcome statistics (and collapsed state) as the proper (so to speak) first-quantized description of the measurement. It's a miracle of living one's life in the symmetric subspace. 

In the real world we are fundamentally forbidden from accessing photons independently in a first-quantized picture. But there is nothing preventing the parties $R,S,T$ who hold the third-quantized state of Eq.~(\ref{Sigma3}) from doing so. In fact their measurement of something like $\{\ket{a}\bra{a},I-\ket{a}\bra{a}\}$ is simply to do photodetection on the first of the three modes they hold, leaving the other two untouched. Depending on the interferometer applied before measurement, some number $n$ of the parties will see the $\ket{a}$ outcome, which would correspond in the original protocol they are simulating to detection of $n$ photons in mode $a$, i.e. projection onto $\fket{n_a}$.   

The upshot of all this is that three parties who hold a state $\ket{\Sigma_{RST}}$ can simulate an arbitrary protocol that someone might want to implement on three single photons, including measurements performed on subsets of modes and feedforward (acting outcome-dependent interferometers on unmeasured modes before they in turn are measured). 

From Knill, Laflamme and Milburn \cite{Knill2001} we know that this procedure, generalized to $N$ photons, is universal for quantum computing. For connoisseurs of photonic quantum computing the rest of the story is obvious, but the appendices contain some more details for completeness. (Performing the third-quantized version of \cite{Knill2001} will not be the most resource-efficient and/or robust use of this type of entanglement, as much more efficient standard photonic quantum computing protocols are known \cite{bartolucci2021fusionbased}. It is also safe to presume there are more intrinsic methods of using the third-quantized entangled state, i.e. not taking the detour through a first-quantized encoded implementation of a standard photonic quantum computation!) 

As with the Bell experiment example we will not know which of the $N$ parties actually have the photons in their lab - the parties do not share the state $\ket{\Sigma}$, rather they can be considered to effectively hold a mixture over every possible way that a subset of $N$ out of the $K$ parties can share the (fully symmetric) state $\ket{\Sigma}$ - the dephased version of Eq.~(\ref{WK3}). But simulating a first-quantized protocol automatically involves all parties doing identical operations - there is no need for anybody to know a-priori whether they are going to be lucky enough to actually be part of the final computation. 

While the protocol outlined in this section involves performing universal quantum computation by starting with an entangled state and performing a sequence of measurements, there are several ways in which it differs from regular one-way quantum computation (1WQC) \cite{Raussendorf2001, Gross2007}:
\begin{itemize}
\item In a 1WQC running a larger algorithm involves increasing the number of systems but the (small) local dimension of each system remains constant. Here both the number of systems and the local dimension increase (polynomially) with the size of the computation.
\item There are no stabilizer states or operations underpinning the protocol ($|\Sigma\rangle$ is not a stabilizer state, and the interferometers used in the various efficient photonic architectures are not elements of some high dimensional stabilizer group.)
\item All systems undergo identical measurement processes at every step, on most systems a rank$>$1 outcome is obtained which leaves them still entangled with the rest.
\item The classical information of measurement outcomes needs to be announced/processed globally, as every party is involved in every step (until they detect a photon) and needs to adapt their measurements accordingly. 
\end{itemize}

As an aside its interesting to note that the geometric entanglement of $\ket{\Sigma}$ goes as $E_g(\ket{\Sigma})\approx N-O(\log N)$ which violates at least the spirit of the result in \cite{flammiagrosseisert} which would suggest that the state is too entangled for 1WQC.

\section{Practical implications}

In this section we consider ways in which this architecture can improve on what are generally considered fundamental barriers within the standard approaches to photonic quantum computing as well as briefly outline potential implications for \emph{other} models of quantum computing

One set of questions about practical implications of  this model of quantum computing revolve around whether it actually should be pursued, as presented, as a way of building a quantum computer. At first glance it definitely should not. Splitting single photons over so many spatial modes that no two photons go the same place would mean some absurdly large footprint - although the modes used could, for example, be different frequencies in which case the spatial footprint is greatly reduced.  

There are multiple ways in which the architectural requirements as presented thus far are much more stringent than actually necessary. For example: its fine if the sources output $n>1$ photons (as long as the number is known it will be useful) and it is not necessary that the phases $\theta_i$ in every copy of $\ket{W_K}$ be the same (they just need to be known). 

It is also fine if more than 1 photon ends up in the laboratory of a single party. However, once we allow for this possibility we need to be careful as to what the rules of the game are - for example, we could send all single photons to a one party who just performs a standard linear optical quantum computation! Obviously we must exclude protocols where the local Hilbert space dimension used by each party grows exponentially. Less trivially, we can, in fact, recover possibilities much closer to regular 1WQC. That is, we can find a way of distributing photonic entanglement using W-states $\ket{W_K}$ with $K=O(1)$, where the number of modes held locally by each party is also $O(1)$ (ensuring the local Hilbert space dimension does not grow with computation size), and where using only linear optical operations and classical communication a computation can be performed. The protocols are more tricky to describe, and so for both pedagogical purposes and because the physics is more interesting we have focussed on the extremal limit with $K\gg N$.

\subsection{Probabilistic (heralded) sources are just fine} \label{sec:probabilisticsources}

It is generally presumed that photonic quantum computing requires a source of (almost) deterministic single photons. Consider, however, taking $K$ heralded sources and sending their outputs into an interferometer described by an $K\times K$ unitary matrix with $|U_{ij}|^2=1/K$ (a ``complex Hadamard'' matrix), then in situations where 1 and only 1 source has heralded an input the output of the interferometer is a state of the form $\ket{W_K}$. (We could arrange to block extra inputs if more than one source fires). Now on each iteration the phases in the $\ket{W_K}$ would be different (they depend on which source fired, and hence which column of $U$ described the spreading). The $K$ possible single photons states that could be produced are all orthogonal - seemingly violating the condition we only use ``identical'' photons for photonic quantum computing. But note that a local phase shifter can adjust the phases as required and this adjustment could be incorporated later into the linear optical transformations that are a necessary part of the eventual computation. Thus  non-deterministic heralded photons can still generate a universal entanglement resource in the sense described above.

\subsection{Improved preparation of standard dual-rail encoded Bell pairs} \label{sec:bleeding}

Above we described how to use third-quantized states of the form $\ket{\Sigma}$ for quantum computing, where the protocol operated in such a way that every party always implements the same operations on the modes they are holding. But this was because we were mimicking a first-quantized protocol. If we had such an entangled state there is nothing per se forcing this upon us. So we can consider protocols for manipulating the entanglement in this state wherein the various parties perform \emph{different} operations - basically we can do linear optical quantum computing without the constraint the photons are identical (i.e. not constrained to operate only in the fully symmetric subspace)! The first task we turn this observation towards is that of preparation of standard, dual-rail encoded, Bell states.

For regular approaches to photonic quantum computing using a dual-rail encoding of a qubit makes sense because it provides the first level of error detection: photon loss, the dominant error mechanism, evolves the qubit outside of the computational subspace. Bell pairs are the basic stabilizer state from which we can create larger stabilizer states, and thence perform fault tolerant quantum computing, so much attention has focussed on methods for creating dual-rail encoded Bell pairs.  

Using only single photons and linear optics it is possible to create a dual-rail encoded Bell state $\fket{1010}\pm\fket{0101}$ (up to permutations of modes) with probability 1/4 by sending 4 single photons through an 8-mode interferometer, and doing photodetection on 4 of the output modes \cite{Browne2005,Zhang2008,Joo2007}. Despite considerable numerical resources being thrown at the problem, the most sophisticated analysis to date being that of \cite{staj}, no improvement over this approach was found.

Recently the $1/4$  barrier was broken in  \cite{stategenerationpaper}, using an approach termed there ``bleeding'', where it was shown how to create a dual rail Bell state with probability up to $2/3$ from four single photons. Here we give an alternative explanation of how bleeding works using a third-quantized picture:  

Consider if parties $A$, $B$, $C$, $D$ shared the third-quantized state of four photons:
\begin{align}
\ket{\Sigma_{ABCD}}=\frac{1}{\sqrt{4!}}\big(&\fket{1000}_A\fket{0100}_B\fket{0010}_C\fket{0001}_D+\ldots \nonumber \\ \ldots+&\fket{0001}_A\fket{0010}_B\fket{0100}_C\fket{1000}_D\big). \nonumber\\ 
  \end{align} 
This state is highly entangled. In fact it can be written in the form
\[
\ket{\Sigma_{ABCD}}:= \frac{1}{\sqrt{6}}\sum_{i=1}^6 (-1)^i\ket{B_i}_{AB}\ket{B_i}_{CD}
\]
where, up to permutation of modes, the states $\ket{B_i}$ are dual rail Bell states of the form $\fket{1010}\pm\fket{0101}$.

As mentioned, there is no reason all parties need perform the same operations given a third-quantized state. Consider then that $A$ and $B$ each send their four modes through an interferometer described by $H\otimes H$ (where $H$ is the Hadamard matrix) and then do photodetection on all four modes. The will each have performed a projection onto four orthogonal states of the form
\begin{equation}\label{eq:HHmeasurement}
\fket{1000}\pm\fket{0100}\pm\fket{0010}\pm\fket{0001}
\end{equation}
A simple calculation shows that with probability $1/2$ they will obtain outcomes that collapse $C$ ad $D$ to holding a Bell state. With probability $1/2$ they collapse $C$ and $D$ to holding a state which is a superposition of all ways of placing two photons in four modes, i.e. of the form
$$
\fket{1100}+\fket{0011}\pm(\fket{1010}+\fket{0101})\pm(\fket{1001}+\fket{0110}).
$$
Such states can be converted (via linear optics and another measurement) into a dual rail Bell state with probability 1/3 \cite{Joo2007,stategenerationpaper}.

The conclusion then is that given a 4 photon entangled state $\ket{\Sigma}$ we can create a desirable Bell state with probability $1/2+1/2\times 1/3=2/3$.

Of course we cannot easily create $\ket{\Sigma}$ from four single photons (as far as we know) but we \emph{can} create $\ket{W_K}^{\otimes 4}$ that we have seen above can be interpreted as ``a random subset of four of the $K$ parties are holding the state $\ket{\Sigma}$''. The problem with this, however, is that we do not know which four of the $K$ parties ``really have'' the photons, and the method just described for generation of a Bell state from $\ket{\Sigma}$ required two of the four parties to do a measurement while the other two should not.
 
 The solution to this conundrum is to proceed sequentially through the $K$ parties \emph{until} two of them have detected a photon (i.e report successful measurement in their local bases of the form Eq.~(\ref{eq:HHmeasurement}). The Bell pair created is delocalized across many modes (since we are not sure which of the unmeasured parties will end up as the $C$ and $D$ in the protocol. But with a little (not particularly enlightening) messing around we can show it is possible to re-localize the qubits. In \cite{stategenerationpaper} a much simpler and more efficient variant of this protocol that basically spreads photons into time instead of space is presented.\footnote{Given a state of the form $\ket{\Sigma}$ one could contemplate two parties performing joint measurements on their systems. But given states of the form $\ket{W_K}^{\otimes N}$ this is not possible, since we do not know which parties ``really'' hold the photons until after they happen to be detected.} 

The bigger picture view here is that we can extend to linear optical quantum information processing certain processes that we might have considered fundamentally forbidden by particle indistinguishability.

\subsection{Universality of W-state preparation and measurement}\label{sec:Wuniversal}

It is easy to think of W-states as the inferior cousins of stabilizer states, particularly when it comes to quantum computing for  which the latter are essential to achieving practical fault tolerance. In certain many-body, matter-based systems (e.g. certain spin-chains, or systems with permutation invariance) W-states (or Dicke states) arise naturally as energy-eigenstates or meta-stable states. As such it still seems worthwhile to understand abstractly what they are or are not useful for. 

We have seen above that in terms of the distribution of entanglement we can use W-states as a resource for quantum computing (including a fault-tolerant version thereof, which it can inherit from the fault tolerance of the underlying standard photonic quantum computing protocol it is simulating). However, the protocol for using those W-states involved the parties continually performing POVMS on subsets of the systems (modes) in their possession. How should we decide if the POVMs are sufficiently W-like that we can claim a completely W-state based protocol for quantum computing?  

One possible criterion could be that all elements of every $n$-mode POVM performed should have a non-degenerate spectral decomposition of the form 
\begin{equation}
\sum_{\alpha=0}^{m<\infty} \lambda_\alpha W^{(\alpha)}
\end{equation}
where $W^{(0)}$ is the projector onto vacuum and the remaining $W^{(\alpha)}$ 
are projectors onto W-states of the form in Eq.~(\ref{eq:Wstate}), with $K=n$, and with phases $\theta^{(\alpha)}_j$ chosen such that they are orthogonal.

It turns out that running the third-quantized version of KLM \cite{Knill2001} does not result in POVMs that satisfy such a criterion. However, running a protocol that involves generation of Bell states via the scheme in \cite{Zhang2008}, and the fusion operations of \cite{Browne2005}, will meet the above criterion.

This is not fully satisfactory, because the third-quantized protocols involve subtle uses of collapse that are not present in standard approaches to measurement-based quantum computing (wherein the systems are typically measured completely - i.e in such a way that the POVM element $E$ determines the collapse rule). Here the collapse rule (which typically involve reduction of Hilbert-space dimension, because interferometers are performed and then only a subset of modes actually detected) is determined by the Kraus operator $K$ (with $K^\dagger K=E$). The natural way to define ``W-like'' Kraus operators is less obviously constrained and should probably be motivated by details of physical systems of interest.

Caveats aside, it seems clear that W-state type of entanglement is interesting and useful for quantum computing and this topic deserves deeper investigation.

\subsection{Boson sampling and the tensor permanent}\label{sec:beyondperms}

For its difficulty Boson sampling \cite{Aaronson2011b} relies on the fact that the amplitudes for ballistically scattering photons through an interferometer and then detecting all modes are functions of the permanent of the unitary matrix which describes the interferometer. Interestingly, an algorithm of Gurvits \cite{Gurvits:2005aa} for finding an additive approximation to the value of a matrix permanent for any matrix with operator norm less than 1, implies that running a Boson sampling experiment cannot be useful in the sense of letting us estimate matrix permanents better than a classical algorithm.    

What we learn from this is that adaptivity is (likely) crucial to performing universal photonic quantum computing: i.e. the interferometers we perform on unmeasured modes necessarily depend on the outcomes of previous measurements.

We now contemplate an ``intermediately adaptive'' type of protocol. We start with $\ket{W_K}^{\otimes N}$, and the first party applies an interferometer $U^{(1)}$ to all $N$ of their modes and measures them. It is likely they do not observe a photon, because $K \gg N$. So the second party attempts to do the same thing, using the same interferometer $U^{(1)}$. This continues until eventually one party detects a photon, say in mode $k_1$. The process continues, however now all subsequent parties use a different interferometer $U^{(2)}$ that need bear no relation to $U^{(1)}$. Eventually a second photon is detected, in mode $k_2$, and once again all subsequent parties switch to using yet another interferometer $U^{(3)}$. This goes on until all $N$ initial photons are detected.

To hopefully achieve an intermediate-strength of adaptivity we fix the interferometers $U^{(1)},\ldots,U^{(N)}$ a-priori. That is, unlike the adaptivity which yields universal quantum computing, $U^{(n)}$ will \emph{not} depend on $k_1,\ldots,k_{n-1}$. Also, here each party measures all the modes they hold when it is their turn - unlike the method for achieving universality via the third-quantized implementation of a regular photonic quantum computation described above. Both limitations make practical implementation considerably easier, and suggest this protocol is not going to be trivially BQP-complete.    

In order to see why this protocol may be harder than standard Boson sampling, we note that the probability of the observed set of outcomes is $P(k_1,k_2,\ldots,k_N)=|Per(M)|^2$ where 
\begin{equation}
M=\frac{1}{\left({N!}\right)^\frac{1}{2N}}\left[\begin{matrix}\bra{U^{(1)}_{k_1,:}} \\ \bra{U^{(2)}_{k_2,:}} \\ \vdots \\ \bra{U^{(N)}_{k_N,:}}\end{matrix}\right]
\end{equation}
Here $\bra{U^{(n)}_{k_1,:}}$ denotes the $k_1$'th row of $U^{(n)}$.

The matrix $M$ can have operator norm greater than 1. Gurvits' algorithm can be extended to matrices with repeated rows \cite{aaronson2012generalizing} (equivalent to detection of multiple photons in a single mode), and such matrices can also have an operator norm larger than 1, but here the $U^{(n)}$ need have no relation to each other. $Per(M)$ can be written as a ``slice'' of a well known tensor generalization of the permanent \cite{wang2018permanent,Dow:1987aa,Taranenko:2016aa}. Gurvits' algorithm for the matrix permanent does not obviously extend to the tensorial generalization \cite{gurvitsprivatecomm}. The tensorial generalization is a significantly more complicated object than its matrix counterpart (e.g. even deciding whether it is 0 or not given a non-negative tensor is NP hard).\footnote{There is a simple quantum algorithm for sampling e.g a 3d-permanent given a standard universal quantum computer - prepare a suitable symmetrized state $\ket{\Sigma}$ and apply $U_1\otimes U_2\otimes \ldots..$ to each qudit, then measure. The complexity of such sampling is perhaps of independent interest.} 

Thus, it seems plausible that this ``intermediately adaptive boson sampling'' problem is more challenging for classical algorithms than the regular version, without being BQP complete.

 \section{Conclusions}

In matter-based systems creation the of useful entanglement involves interaction. For photons it does not - free evolution generates considerable amounts of entanglement deterministically. Photons can generate high dimensional entanglement both by evolving into more modes, and by inhabiting higher occupation numbers within a single mode. (The latter possibility has not been considered here.) Yet photonic quantum computing typically follows routes originally conceived for matter-based systems. These are unnatural and inefficient. It takes considerable effort to free photons from the qubit tyranny - the  obstacle is simply that fault tolerance is very much better understood for stabilizer states, and the qubit versions thereof in particular.

It is generally believed (by the physicists on this planet) that to claim a true understanding of quantum theory one of our tasks is an explanation of why useful entanglement is so fragile that the world we experience is classical. Entanglement, in our experience, only manifests itself when the cleverest of our species capture and protect it appropriately in controlled and delicate experiments. However, once one comprehends how incredibly robust, pervasive and useful photonic entanglement is the question is flipped - why is it that we did not evolve to make use of it? Why is it that this entanglement did not help us in finding mates or bananas?

\acknowledgements
I am grateful to the Architecture team at PsiQuantum for our many explorations in photonic quantum information together. In particular I would like to thank Eric Johnston for the linear-optical simulations software tool, without which my understanding of photons would be much poorer.

\bibliography{LOQCpapers3}

\appendix*


\section{FIRST QUANTIZATION, FAUX QUANTIZATION AND PHOTONIC QUANTUM COMPUTING} 

In this appendix we spell out in a little more detail the result of the main section.

\subsection{Second-quantized description of standard photonic quantum computing}

When we talk about photonic quantum computing we normally describe things in the second-quantized picture, where the systems are modes, and photons are an internal state (excitation) of the system. With this language all approaches to photonic quantum computing that use a finite number of photons initially in Fock states can be described in the following general terms:
\begin{enumerate}
	\item The initial state is a product state across $M$ modes. While each of these $M$ systems is formally infinite dimensional (i.e. could be occupied by an arbitrary number of photons), in fact we begin with a fixed total number $N$ of (typically single) photons, and so they need be considered at most $N$-dimensional (i.e. ``qu-$N$-its'') 
	\item[2a] Subsets of the modes are combined at interferometers, which is a multi-system unitary evolution. Entanglement is created by the interferometers. Then a fraction of the output modes from each interferometer are detected (i.e. there is a \emph{partial} measurement, effectively a POVM is performed).  This photodetection \emph{is not} a joint (i.e. entangled) measurement.
	\item[2b] Based on the detection patterns in Step 2a, and some side classical computation that also depends on the algorithm being run, other subsets of modes are chosen to undergo a POVM similar to that in Step 2a.
	\item[2c]...
	\item[3] Eventually all modes are measured, the output of the computation is efficiently extractable from the classical measurement record.    
\end{enumerate}

\subsection{First-quantized description of standard photonic quantum computing}

 In the first-quantized picture the systems are photons and the modes each photon can occupy are its internal states. In this language the exact same computation sounds somewhat different: 
\begin{enumerate}
	\item The initial state is an \emph{entangled} state across $N$ photons. While each of the $N$ systems is formally infinite dimensional (i.e. could occupy an arbitrary number of modes), in fact we begin with a fixed total number $M$ of modes, and so they need be considered at most $M$-dimensional (i.e. ``qu-$M$-its'') 
	\item[2a] Subsets of the modes are combined at interferometers, which induce an independent (but identical) unitary evolution of each system. Entanglement is \emph{not} created by the interferometers. Then a fraction of the output modes from each interferometer are detected (i.e. there is a \emph{partial} measurement, effectively a POVM performed).  This photodetection \emph{is} a joint (i.e. entangled) measurement.
	\item[2b] Based on the detection patterns in Step 2a, and some side classical computation that also depends on the algorithm being run, other subsets of modes are chosen to undergo a POVM similar to that in Step 2a.
	\item[2c]...
	\item[3] Eventually all modes are measured, the output of the computation is efficiently extractable from the classical measurement record.    
\end{enumerate}

In the first-quantized description preparation/measurement of number states of photons is preparation/measurement of highly entangled (i.e. the fully symmetrized) states. An interferometer is described by some $M$-dimensional unitary $U$ that acts independently as the $N$-fold tensor product $U\otimes U\otimes \ldots\otimes U $ across the $N$ systems, and so is not entangling. The role of entanglement in describing how the computation works differs greatly between first and second-quantized descriptions although they are formally isomorphic.  

Superficially this comparison is somewhat trivial - take any quantum computing architecture, refactorize the Hilbert space, and you likely will have some complicated description involving entanglement where perhaps there was none before.

\subsection{Faux-quantized description of standard photonic quantum computing}

In \cite{popescuKLM} Popescu pointed out that within the first-quantized description there is no need for the measurements (projections onto Fock states) to be of the joint (entangling) type that the rigorous mapping between first and second quantization dictates. The highly symmetric structure ensures that we can actually replace the single joint projective measurement (of the first-quantized description) with an $N$-fold tensor product of identical, but independently performed, projective measurements! (An example is given below). Using real physical photons, nature prevents us from accessing them independently so as to measure them separately - so this model of quantum computing is no longer isomorphic to the first-quantized description of a protocol. Yet it is ``just as good'' in as much as it has the same ability to efficiently simulate any photonic quantum computing protocol and therefore perform a universal quantum computation. 

This ``distinguishable first-quantized photons'' model begins to look a little more like the regular one-way quantum computer \cite{Raussendorf2001} - individual system measurements on an initially entangled state drive universal quantum computation.
As such, lets imagine that we implement Popescu's proposal by taking $N$ standard qu-$M$-its for which we are not constrained by the physical laws of particle indistinguishability, and further imagine we have created the same fully-symmetrized entangled state $\ket{\Sigma}$ that $N$ photons in $M$ modes are described by in first quantization. We will call this qu-$M$-it protocol that mimics first quantization (but with distinguishable systems and independent system measurement) the \emph{faux-quantized} picture.   

To make things a bit more concrete, let us label modes $a,b,c,\ldots$ such that $\ket{a_j}$ denotes photon/qu-$M$-it number $j$ in mode/level $a$ for the first/faux-quantized description. As an example, for three single photons/qu-$M$-its we have
\begin{align*}
\ket{\Sigma}&=\frac{1}{\sqrt{3!}}\left(\ket{a_1}\ket{b_2}\ket{c_3}+\ket{a_1}\ket{c_2}\ket{b_3}+\ldots\ket{c_1}\ket{b_2}\ket{a_3}\right) \\
  &\Leftrightarrow \fket{1_a}\fket{1_b}\fket{1_c}
\end{align*}
where on the second line the corresponding Fock state in second quantization is shown explicitly ($\fket{\cdot}$ brackets always denote a second-quantized state of physical photons).

Consider some protocol dictating modes $a$ and $b$ pass through an interferometer $V\in SU(2)$ and then a detector is placed in mode $a$ with a single photon being detected. In the faux protocol we evolve the state as we would in first-quantization:
\[
\ket{\Sigma'}=\left(\begin{bmatrix}V &0 \\ 0 & 1 \end{bmatrix}\otimes \begin{bmatrix}V &0 \\ 0 & 1 \end{bmatrix} \otimes \begin{bmatrix}V &0 \\ 0 & 1 \end{bmatrix}\right) \ket{\Sigma}.
\]

However, rather than the joint measurement over all three systems of first quantization, in the faux protocol the detection in mode $a$ is now mimicked by performing three independent measurements of the pair of projectors $\Pi_{a_j}= \ket{a_j}\bra{a_j}$, $\Pi_{\bar{a}_j} =\mathbb{I}-\Pi_{a_j}$. Detection of a single photon means on one qu-$M$-it the $\Pi_a$ outcome obtains, on the other two the $\Pi_{\bar{a}}$ outcome pertains. 

Note that the measurement performed on each qu-$M$-it is identical, and the entanglement of the state is not (typically) fully destroyed because of the high rank $\Pi_{\bar{a}}$ types of outcomes. The next step of the protocol would involve a similar process - identical unitaries on each system, followed by another projective measurement. This can be thought of as simply a POVM that acts nontrivially on a subspace of each systems (large) Hilbert space.  In fact, normally a photonic computation starts with $N$ single photons and $M>N$ modes (i.e. vacuum ancillary modes are necessary). From the faux implementation perspective, the role of the vacua is only to let us do a generic (high rank) POVM using a ``direct sum ancillary subspace'' rather rather than the more typically encountered ``tensor product ancillary system''. That is, if we have the ability to do arbitrary POVMs then we only only require qu-$N$-its not qu-$M$-its to run the faux version. (As an aside, the fact that we do have some protocols which are finite depth with respect to an individual photon's worldline implies some interesting variants with even more constrained local dimensions are possible, this is a topic for elsewhere.)

In summary, the faux version of any protocol begins with $N$ distinguishable qu-$M$-its whose internal levels are the ``modes'' of the protocol. The faux-quantized description of mimicking a photonic quantum computation can be summarized:
\begin{enumerate}
	\item The initial state is an \emph{entangled} state $|\Sigma\rangle$ across $N$ qu-$M$-it systems.  
	\item[2a] Subspaces of each qu-$M$-it's internal levels are identically and independently nontrivially rotated (equivalent to subsets of the modes evolving through interferometers). Entanglement is \emph{not} created by the unitary evolution. Then a \emph{partial} measurement, effectively a POVM, is performed separately on each qu-$M$-it (equivalent to a fraction of the output modes from each interferometer being detected).  This \emph{is not} a joint measurement.
	\item[2b] Based on the detection patterns in Step 2a, and some side classical computation that also depends on the algorithm being run, other subspaces of each qu-$M$-it (subsets of modes) are chosen to undergo a POVM similar to that in Step 2a.
	\item[2c]...
	\item[3] Eventually all qu-$M$-its are fully measured, the output of the computation is efficiently extractable from the classical measurement record.    
\end{enumerate}

This general faux-quantized picture is completely agnostic about whether we are doing a KLM version of photonic quantum computing, a cluster state version \cite{nielsenloqc,Browne2005}, Fusion Based Quantum Computing \cite{bartolucci2021fusionbased}, ... or hybrids thereof. Those distinctions change only the specific POVMs we do on each system. While it is has similarities with standard measurement-based approaches, as itemized at the end of Section~\ref{sec:gettingtofull} there are also many differences. The primary limitation of this model, practically speaking, is that there are no simple and robust preparation methods for $\ket{\Sigma}$.

\subsection{Third quantization to do a photonic implementation of fauxtonic quantum computing}\label{thirdquantization}

If there was some way to easily create the highly entangled state $\ket{\Sigma}$ using $N$ distinguishable qu-$M$-its then it would be worth exploring whether faux-quantization opened some new practical route to quantum computing. At present no such result is known. 
We will see now, however, that there is a simple procedure for creating a state that is closely related to $\ket{\Sigma}$ and that inherits its usefulness for a faux-quantized implementation of quantum computing. Perversely, this procedure is particularly easy to implement with photons, which can lead to some confusion since we are using our fundamentally indistinguishable particles to encode a state over systems that we want to be able to measure independently, i.e. as if they were distinguishable! We will call this perverse encoding the \emph{third-quantized} state. 

We begin by considering how to encode qu-$M$-its with photons. The natural way is to use a single photon in $M$ modes:
\begin{equation*}
\ket{a}=\fket{100\ldots0}\,,\ket{b}=\fket{010\ldots0}\,,\ldots
\end{equation*}
On such states completely arbitrary unitary evolution and arbitrary POVMs are easily performed by interferometers and photodetection. Note that, although there is only one photon present, a general superposition of such states is entangled across modes, and so such operations are non-trivial - they require generation and manipulation of entanglement between the modes. 

Revisiting the example of $\ket{\Sigma}$ from above, the third-quantized encoding of this state has three photons in nine modes and takes the form
\begin{align}\label{sigma3}
\ket{\Sigma}=&\frac{1}{\sqrt{3!}}\left(\ket{a_1}\ket{b_2}\ket{c_3}+\ket{a_1}\ket{c_2}\ket{b_3}+\ldots+\ket{c_1}\ket{b_2}\ket{a_3}\right) \nonumber\\ 
  = &\frac{1}{\sqrt{3!}}(\fket{100}\fket{010}\fket{001}+\fket{100}\fket{001}\fket{010} \nonumber\\
   &+\ldots +\fket{001}\fket{010}\fket{100})
\end{align}
(If $M>N$ then we can just add on some vacuum modes to the end of each photonic qu-$M$-it.)

If we had such a state then implementing the faux-quantized version of a photonic protocol would be easy. Interestingly it would not involve any kind of ``truly bosonic'' behaviour, in as much as there would never be  states with more than one photon in a mode (such as $\fket{2}$, $\fket{3}$ etc) created.

It seems likely that, just as with preparing photonic versions of entangled stabilizer qubit states, preparation of the third-quantized  $\ket{\Sigma}$ would require tricky probabilistic procedures involving multiple photon interference.  We will now see, however, that starting with single photons we can easily and deterministically create a different state that is equivalent to $\ket{\Sigma}$ from the perspective of implementing any \emph{fauxtonic} protocol. In fact creating this new state does not require any interference of multiple photons at all, nor probabilistic gates. 

Going back to the simple third-quantized example above, imagine that each of the 3 qu-$M$-its in the example is held by a different party $A, B,C$. We could add three more persons, $D,E,F$ and give them 3 empty modes each and the protocol could remain unchanged. Consider then the mixed state
\begin{equation*}
\frac{1}{2}\ket{\Sigma_{ABC}}\bra{\Sigma_{ABC}}+\frac{1}{2}\ket{\Sigma_{DEF}}\bra{\Sigma_{DEF}}.
\end{equation*}
Here $\ket{\Sigma_{ABC}}$ indicates that it is the first three parties who actually have a single photon (and so share the entanglement of $\ket{\Sigma}$) while $\ket{\Sigma_{DEF}}$ indicates it is the latter three.   

We could implement any faux-quantized protocol perfectly using the above mixed state, because each and every party in the faux version of a protocol performs identical actions to one another. The fact that it is only at the end that some parties will discover that they actually ``really took part'' (the rest having always detected vacuum) makes no difference to how the computation proceeds. In fact a mixture containing states that have some parties in common (such as a mixture of $\ket{\Sigma_{ABC}}$ and $\ket{\Sigma_{ACD}}$) is similarly fine. 

More generally then, if we consider a case where we have $K\ge N$ parties, and let $\alpha$ denumerate all $N$-fold subsets of parties, then we see that the mixed state 
\begin{equation*}
\rho(K,N,M)=\frac{1}{\binom{K}{N}}\sum_\alpha \ket{\Sigma_\alpha}\bra{\Sigma_\alpha}
\end{equation*}
can be used to do the faux version of any photonic protocol. In fact with a little thought we see that a mixture is not necessary, the pure state 
\begin{equation*}
\ket{\Sigma^\star(K,N,M)}:=\frac{1}{\sqrt{\binom{K}{N}}}\sum_\alpha \ket{\Sigma_\alpha}
\end{equation*}
would also be fine (each party could apply a random phase to all the modes they hold  prior to the computation - this would commute through the rest of what they do in the protocol).
      
Consider the case where we take a single photon and spread it uniformly over $K>>N$ modes; denote the resultant state $|W_K\rangle$. In second-quantized form we can write
\begin{align*}
\ket{W_K}&=\frac{1}{\sqrt{K}}\sum_{j=1}^K\fket{1_j} \\
&=\frac{1}{\sqrt{K}}(\fket{100..0}+\fket{010..0}+\ldots\fket{00..1}).
\end{align*}
The first mode will go to party $A$, the second to party $B$ and so on. Repeating this whole process with $N$ copies of $|W_K\rangle$, each party will now hold $N$ modes. (If $M>N$ we can give them additional vacuum modes.)

By choosing $K$ large enough the probability that any party holds more than one photon can be reduced as small as we like. (The majority of parties will actually hold 0 photons.) Moreover, if the $k$'th party holds the $j$'th photon, then no other party will be holding it. If we restrict attention in the total state to a fixed subset of the parties who do have a photon, we see the state is a superposition of all $N!$ distinct ways they can hold the photons in different modes - just as in the example of Eq.~(\ref{sigma3}) above.   We readily deduce that the state $|W_K\rangle^{\otimes N}$ is approximately a third-quantized state:  
\begin{equation*}
|W_K\rangle^{\otimes N}\approx \ket{\Sigma^\star(K,N,M)}
\end{equation*}
(e.g. taking $K=N^3$ the error in the approximation is $O(1/N)$).
In summary, we can take any photonic approach to quantum computing, and implement it as a faux-quantized style protocol in a third-quantized manner that never involves ``genuine'' multiphoton interference. In the third-quantized approach what varies as we change which regular photonic quantum computing protocol we are simulating is only the details of the (single photon) POVM that each party performs. While not particularly practical, conceptually this is pretty interesting. Note that since we never bring multiphoton terms into the picture, we could map the photonic $\fket{0}$ and $\fket{1}$ to a generic qubit state, and so we can potentially use these ideas more abstractly to say things about the usefulness of things like $W$-states for regular qubit-based quantum computing.

%

\end{document}